\documentclass{article}
\usepackage[utf8]{inputenc}

\usepackage{amsmath}
\usepackage{amssymb}
\usepackage{hyperref}
\usepackage{graphicx}
\usepackage{verbatim}
\usepackage{geometry}
\usepackage{slashed}
\usepackage{tikz}
\tikzstyle{every picture}=[level distance = 8mm, baseline=-0.5ex]
\tikzstyle{prop}=[shape=circle,minimum size=6mm, draw=black!80, fill=green!30]
\tikzstyle{propScal}=[shape=rectangle,minimum size=6mm, draw=black!80, fill=green!30]

\usetikzlibrary{arrows}

\renewcommand{\d}{\text{d}}

\newcommand{\E}{\text{Erfi}}

\begin{document}

\title{Analytic results for Schwinger--Dyson equations with a mass term.}
\author{Pierre~J.~Clavier${}^{1,2}$\\
\normalsize \it ${}^1$ Sorbonne Universités, UPMC Univ Paris 06, UMR 7589, LPTHE, 75005, Paris, France\\
\normalsize \it $^2$ CNRS, UMR 7589, LPTHE, 75005, Paris, France }

\date{}

\maketitle

\begin{abstract}

Using kinematic renormalization, we derive the Schwinger--Dyson equations for a massive Yukawa model and a Wess--Zumino-like one. Both have linear Schwinger--Dyson equations and a massive 
renormalized particle. An explicit solution is found in the IR limit of the non-supersymmetric case. Parametric solutions are found in the UV limit of the same model and for the supersymmertic 
model.

\end{abstract}

\section*{Introduction}

As a way to reach non-perturbative informations of a Quantum Field Theory (QFT), the Schwinger--Dyson equations have been quite extensively studied over the past few decades. The most common 
approach to tackle them was, and maybe still is, numerical study of simplified versions of the equations, that however display important physical features, e.g. of QCD. For a recent paper in this domain,
the reader can referred to \cite{MeMoRoSa14}.

On the other hand, some progresses have been made these last few years to analytically study some Schwinger--Dyson equations. The anomalous dimension (from which the dressed propagator could be 
extracted using the renormalization group equation, as shown in \cite{KrYe2006}) have been computed order by order in \cite{BeSc12}. Its asymptotic has also been extracted in the same article. 
Corrections to this asymptotic behaviour have been computed in \cite{BeCl13}. This have been rephrased in the Borel plane in \cite{BeCl14}, where some number-theoretical results could be shown 
to hold at every order in the corrections to the asymptotic behaviour of the anomalous dimension.

However, there is still an important lack of exact known solutions of Schwinger--Dyson equations. To our knowledge, the only one is for a massless and linear equation and was found by 
David Broadhurst and Dirk Kreimer in \cite{BrKr99}. This paper aims to be a step toward filling this hole in the field of Schwinger--Dyson equations by applying the method of \cite{BrKr99} to more 
general cases.

The result of \cite{BrKr99} rests upon three crucial steps. First, the integrodifferential Schwinger--Dyson equation is written as a differential one. This equation, when written with the right functions, 
can be integrated in the second step. Then, as a third step, it can be solved for some functions having the integrated functions as arguments. Using the initial conditions, we end up with a 
parametric solution of the initial Schwinger--Dyson equation.

The first of this steps is made possible by the linearity of the initial equation. We therefore will only study here linear equations. In order to perform the angular integration in the loop 
integrals it is needed to have the field without corrections to its two points function to be massless. Here, we will keep that constraint. However, the second and third steps can 
give results for theories more general than the one studied in \cite{BrKr99}. In some cases (infrared limit of massive Yukawa model) we end up with equations simple enough to be 
solve explicitly (although only at a given impulsion of reference in the soft IR case).

This paper is organized as follow: in the sections 1 and 2, we derive the Schwinger--Dyson equations for the massive Yukawa model and for a linear version of a Wess--Zumino model (a massive 
version of the model studied in \cite{BeLoSc07}). In both cases, we use kinematic renormalization to do so. In section 3, a parametric solution to the massive Yukawa model in the ultraviolet 
limit is found. Section 4 is devoted to the study of the massive Yukawa model in the infrared limit. Finally, in section 5, a parametric solution to the massive supersymmetric model is found. In 
order to keep the length of this article within reasonable size, a full presentation of the method of \cite{BrKr99} is not included. Nevertheless, a short summary of this method could be found 
in section 3, leaving aside any technical details.

\section{Schwinger-Dyson equations of the massive Yukawa model}

For the Yukawa models, the Schwinger-Dyson equations are linear:
\begin{equation} \label{SDlinnSUSY}
 \left(
\tikz \node[prop]{} child[grow=east] child[grow=west];
\right)^{-1} = 1 - a \;\;
\begin{tikzpicture}
\draw (-1.2,0)--(-0.8,0);
\draw (0.8,0)--(1.2,0);
\draw[dashed] (-0.8,0) .. controls (-0.8,1) and (0.8,1) .. (0.8,0);
\node at (0,0) [circle,minimum size=6mm,draw,fill=green!30] {} child [grow=east] child[grow=west];
\end{tikzpicture}
\end{equation}
with $1$ denoting the free propagator and the dressed fermionic propagator being
\begin{equation}
 P_{nSUSY} = \frac{1}{G(q^2)\slashed{q}+M(q^2)m}.
\end{equation}
In order to find back the free propagator at a the impulsion of reference $\mu$, we have the initial conditions $G(\mu^2)=M(\mu^2)=1$. So the equation (\ref{SDlinnSUSY}) can be written in term of 
Feynman integrals:
\begin{equation}
 G(p^2)\slashed{p}+M(p^2)m = \slashed{p} + m -\frac{2a}{\pi^2}\int\d^4l\frac{G(l^2)\slashed{l}-mM(l^2)}{G^2(l^2)l^2-M^2(l^2)m^2}\frac{1}{(p+l)^2} + S_1\slashed{p}+S_2m.
\end{equation}
$S_1$ and $S_2$ are two counterterms that will be fixed using the initial conditions. Note that we are here using the sign convention of \cite{Itzykson}. The basic idea is that the 
$\slashed{p}$ and the $m$ parts do not talk to each other, so their coefficients should independently vanish. Hence we are left with a system of two coupled integral equations:
\begin{subequations}
 \begin{align}
  & G(p^2) = 1 - \frac{2a}{\pi^2}\int\text{d}^4l\frac{G(l^2)}{G^2(l^2)l^2-m^2M^2(l^2)}\frac{1}{p^2}\frac{l.p}{(p+l)^2}+S_1 \label{SDEnSUSY1a} \\
  & M(p^2) = 1 + \frac{2a}{\pi^2}\int\text{d}^4l\frac{M(l^2)}{G^2(l^2)l^2-m^2M^2(l^2)}\frac{1}{(p+l)^2} + S_2. \label{SDEnSUSY1b}
\end{align}
\end{subequations}
From now on those two equations will be the ones referred as the Schwinger--Dyson equations. The two integrals can be computed by separating their radial and angular parts. Then, using the 
4-dimensional angular average
\begin{equation} \label{angint}
 q^2l^2\left\langle\frac{1}{(q+l)^2}\right\rangle_{d=4} = \text{min}(q^2,l^2)
\end{equation}
we are left with:
\begin{subequations}
 \begin{align}
  & G(p^2) = 1 + \frac{a}{p^2}\int_0^p\text{d}l\frac{l^3G(l^2)}{G^2(l^2)l^2 - m^2M^2(l^2)}\frac{l^2}{p^2} + a\int_{p^2}^{+\infty}\text{d}l\frac{lG(l^2)}{G^2(l^2)l^2-mM^2(l^2)} + S_1 \\
  & M(p^2) = 1 + \frac{2a}{p^2}\int_0^p\text{d}l\frac{l^3M(l^2)}{G^2(l^2)l^2-m^2M^2(l^2)} + 2a\int_p^{+\infty}\text{d}l\frac{lM(l^2)}{G^2(l^2)l^2-m^2M^2(l^2)} + S_2.
\end{align}
\end{subequations}
Let us notice that, for the equations of $G$, we have used $l.p=\frac{1}{2}\left[(p+l)-p^2-l^2\right]$. Then, three terms have combined themselves together to cancel and we are left with such a 
simple equation for $G$. Now, using the initial conditions $G(\mu^2)=M(\mu^2)=1$ allows to fix the counterterms:
\begin{align*}
 & S_1 = -\frac{a}{\mu^2}\left[\int_0^{\mu}\d l\frac{l^3G(l^2)}{G^2(l^2)l^2-m^2M^2(l^2)}\frac{l^2}{\mu^2} + \mu^2\int_{\mu}^{+\infty}\d l\frac{lG(l^2)}{G^2(l^2)l^2-m^2M^2(l^2)}\right] \\
 & S_2 = -\frac{2a}{\mu^2}\left[\int_0^{\mu}\d l\frac{l^3M(l^2)}{G^2(l^2)l^2-m^2M^2(l^2)} + \mu^2\int_{\mu}^{+\infty}\d l\frac{lM(l^2)}{G^2(l^2)l^2-m^2M^2(l^2)}\right].
\end{align*}
Plugging those two counterterms into the Schwinger-Dyson equations, rewriting them with $x=p^2$ and switching the integration variable to $y=l^2$ allow us to write the Schwinger-Dyson equations in
the following compact form
\begin{subequations}
 \begin{align}
  & G(x) = 1 - \frac{a}{2}\int_{\mu^2}^x\d y\frac{G(y)}{G^2(y)y-m^2M^2(y)}+F(x)-F(\mu^2) \\
  & M(x) = 1 - a\int_{\mu^2}^x\d y\frac{M(y)}{G^2(y)y-m^2M^2(y)}+E(x)-E(\mu^2),
\end{align}
\end{subequations}
with:
\begin{subequations}
 \begin{align}
  & F(x) = \frac{a}{2}\int_0^x\d y\frac{G(y)}{G^2(y)y-m^2M^2(y)}\left(\frac{y}{x}\right)^2 \\
  & E(x) = a\int_0^x\d y\frac{M(y)}{G^2(y)y-m^2M^2(y)}\frac{y}{x}.
\end{align}
\end{subequations}
Finally, taking two derivatives with respect to $x$, we can write the Schwinger-Dyson equations as two coupled ordinary differential equations:
\begin{subequations}
 \begin{align}
  & \left[G^2(x)x - m^2M^2(x)\right]D(D+1)M(x) = -axM(x) \label{SDEnSUSY2a} \\
  & \left[G^2(x)x - m^2M^2(x)\right]D(D+2)G(x) = -axG(x) \label{SDEnSUSY2b}
\end{align}
\end{subequations}
With the differential operator $D=x\frac{\d}{\d x}$. Here the key point is that the two integral terms surviving the differentiation process cancel each other. This is why the linearity of the 
Schwinger--Dyson equation (\ref{SDlinnSUSY}) is crucial: it allows to end up with differential equations instead of integrodifferential ones.

Let us notice that in \cite{BeLoSc07} the authors avoid to explicitly write the counterterms by taking the derivatives with respect to $x$ first. This has the advantage that no ill-defined 
integrals appear in the computations. However we choose here to write down the couterterms before taking the derivatives to follow the presentation of \cite{BrKr99}. Obviously the result is the 
same within the two approaches.

One can easily check that taking the massless limit $M(x)=0$ in (\ref{SDEnSUSY2a}) we find the Schwinger-Dyson equation of the massless Yukawa model
already studied in \cite{BrKr99}. Let us notice however that the generalization of the case studied in \cite{BrKr99} is not trivial. In particular, the right-hand-side appears to be now dependent 
of the external momenta.

Now, when decoupling the equations (\ref{SDEnSUSY2a}-\ref{SDEnSUSY2b}), one ends up with complicated non-linear equations having a non-trivial denominator. It makes a rigorous analysis quite 
challenging, since we would have to take care that the denominator does not vanish. Instead, we will rather tackle the equations (\ref{SDEnSUSY2a}-\ref{SDEnSUSY2b}) in the physically 
relevant cases of the ultraviolet and infrared limits. Before we turn our attention to this task, let us derive the Schwinger--Dyson equation of a massive supersymmetric theory.

\section{Schwinger-Dyson equation of the supersymmetric model}

We will work with a massive version of a Wess--Zumino-like model already studied in \cite{BeLoSc07}. This model has two superfields, one massive, $\Psi_i$, and one massless, $\Phi_{ij}$ 
($i,j=1,2,...,N)$. Each superfield represent a complex scalar ($A_i$ or $B_{ij}$), a Weyl fermion ($\chi_i$ or $\xi_{ij}$) and a complex auxiliary field ($F_i$ or $G_{ij}$):
\begin{eqnarray*}
   \Psi_i(y) & = & A_i(y) + \sqrt{2}\theta\chi_i(y) + \theta\theta F_i(y) \\
 \Phi_{ij}(y) & = & B_{ij}(y) + \sqrt{2}\theta\xi_{ij}(y) + \theta\theta G_{ij}(y).
\end{eqnarray*}
With $y$ the chiral coordinates defined by $y^{\mu} = x^{\mu}+i\theta\sigma^{\mu}\bar{\theta}$. This model has a cubic superfield interaction Lagrangian
\begin{eqnarray*}
 L_{\text{int}} & = & \frac{g}{\sqrt{N}}\sum_{i,j=1}^N\int \d^2\theta\Psi_i\Phi_{ij}\Psi_j \\
   & = & \frac{g}{\sqrt{N}}\sum_{i,j=1}^N(A_iG_{ij}A_j + 2A_iB_{ij}F_j - \chi_iB_{ij}\chi_j - \chi_i\xi_{ij}A_j - A_i\xi_{ij}\chi_j + \text{h.c.}).
\end{eqnarray*}
A more detailed presentation of this model can be found in \cite{BeLoSc07}. Such a model is of interest for us due to the existence of non-renormalization theorems, whose simplest proof is 
usually credited to  Seiberg \cite{Se93}. A nice introduction of the subject can be found in \cite{De89}. The theorem implies that one only need wavefunction renormalization, as we will see 
later. The Schwinger--Dyson equations can be graphically written as
\begin{subequations}
 \begin{eqnarray}
  \left(
\tikz \node[prop]{} child[grow=east,densely dotted,thick] child[grow=west,densely dotted,thick];
\right)^{-1} & = & 1 - a \;\;
\begin{tikzpicture}
\draw[densely dotted,thick] (-1.2,0)--(-0.8,0);
\draw[densely dotted,thick] (0.8,0)--(1.2,0);
\draw[dashed] (-0.8,0) .. controls (-0.8,1) and (0.8,1) .. (0.8,0);
\node at (0,0) [circle,minimum size=6mm,draw,fill=green!30] {} child [grow=east,dashed] child[grow=west,dashed];
\end{tikzpicture} \\
  \left(
\tikz \node[prop]{} child[grow=east] child[grow=west];
\right)^{-1} & = & 1 - a \;\;
\begin{tikzpicture}
\draw (-1.2,0)--(-0.8,0);
\draw (0.8,0)--(1.2,0);
\draw[dashed] (-0.8,0) .. controls (-0.8,1) and (0.8,1) .. (0.8,0);
\node at (0,0) [circle,minimum size=6mm,draw,fill=green!30] {} child [grow=east] child[grow=west];
\end{tikzpicture}
-a\;\;
\begin{tikzpicture}
\draw (-1.2,0)--(-0.8,0);
\draw (0.8,0)--(1.2,0);
\draw (-0.8,0) .. controls (-0.8,1) and (0.8,1) .. (0.8,0);
\node at (0,0) [circle,minimum size=6mm,draw,fill=green!30] {} child [grow=east,dashed] child[grow=west,dashed];
\end{tikzpicture} \\
  \left(
\tikz \node[prop]{} child[grow=east,dashed] child[grow=west,dashed];
\right)^{-1} & = & 1 - a \;\;
\begin{tikzpicture}
\draw[dashed] (-1.2,0)--(-0.8,0);
\draw[dashed] (0.8,0)--(1.2,0);
\draw[densely dotted,thick] (-0.8,0) .. controls (-0.8,1) and (0.8,1) .. (0.8,0);
\node at (0,0) [circle,minimum size=6mm,draw,fill=green!30] {} child [grow=east,dashed] child[grow=west,dashed];
\end{tikzpicture}
-a\;\;
\begin{tikzpicture}
\draw[dashed] (-1.2,0)--(-0.8,0);
\draw[dashed] (0.8,0)--(1.2,0);
\draw (-0.8,0) .. controls (-0.8,1) and (0.8,1) .. (0.8,0);
\node at (0,0) [circle,minimum size=6mm,draw,fill=green!30] {} child [grow=east] child[grow=west];
\end{tikzpicture} \\
   \llcorner & - & a \;\;
\begin{tikzpicture}
\draw[dashed] (-1.2,0)--(-0.8,0);
\draw[dashed] (0.8,0)--(1.2,0);
\draw[dashed] (-0.8,0) .. controls (-0.8,1) and (0.8,1) .. (0.8,0);
\node at (0,0) [circle,minimum size=6mm,draw,fill=green!30] {} child [grow=east,densely dotted,thick] child[grow=west,densely dotted,thick];
\end{tikzpicture} \nonumber
 \end{eqnarray}
\end{subequations}
with the plain lines being for the fermionic fields, the dashed lines for the scalar fields and the dotted lines for the auxiliary fields. In the large $N$ limit the one-loop contributions to 
the dressed propagators are the only one to not be suppressed. This is why we consider a model with a vectorial superfield and a matrix one: a solution of the above system is more than a 
solution of a truncated Schwinger--Dyson equation. It is the full dressed propagators of the theory in the large $N$ limit. Now, supersymmetry imposes the following dressed propagators
\begin{subequations}
 \begin{eqnarray}
 \Pi_{\chi}^{-1}(q) & = & \frac{q^2G_{\chi}(q^2)+m^2}{q_m\sigma^m} \\
    \Pi_{A}^{-1}(q) & = & q^2G_{A}(q^2)+m^2 \\
    \Pi_{F}^{-1}(q) & = & \frac{q^2G_{\chi}(q^2)+m^2}{q^2}
\end{eqnarray}
\end{subequations}
where we have dropped the subscript $i$ for simplicity. Hence the system of Schwinger--Dyson equations can be written, after some simplifications as
\begin{subequations}
 \begin{eqnarray}
 q^2G_{\chi}(q^2) & = & q^2 - \frac{g^2}{4\pi^4}\int\d^4p\frac{q.p}{[p^2G_{\chi}(p^2)+m^2](q-p)^2} - \frac{g^2}{4\pi^4}\int\d^4p\frac{q^2-q.p}{[p^2G_{A}(p^2)+m^2](q-p)^2} \qquad ~ \label{SDE_SUSY1} \\
         G_F(q^2) & = & 1 - \frac{g^2}{4\pi^4}\int\d^4p\frac{1}{[p^2G_{A}(p^2)+m^2](q-p)^2} \label{SDE_SUSY2} \\
      q^2G_A(q^2) & = & q^2 - \frac{g^2}{4\pi^4}\int\d^4p\frac{1}{p^2G_{\chi}(p^2)+m^2} - \frac{g^2}{4\pi^4}\int\d^4p\frac{p^2}{[p^2G_{F}(p^2)+m^2](q-p)^2} \label{SDE_SUSY3} \\
        \llcorner & - & \frac{g^2}{4\pi^4}\int\d^4p\frac{\text{Tr}(p_m\sigma^m(q_n-p_n)\sigma^n)}{[p^2G_{\chi}(p^2)+m^2](q-p)^2}. \nonumber
 \end{eqnarray}
\end{subequations}
We did not write explicitly the counter-terms in this system in order to keep it of reasonable size. Now, as already noticed in \cite{BeLoSc07}, a coherent ansatz to solve this system is
\begin{equation}
 G_{\chi}(q^2) = G_F(q^2) = G_A(q^2) = G(q^2).
\end{equation}
Indeed, with this ansatz, the first integral of (\ref{SDE_SUSY1}) cancels the $q.p$ term of the second integral and therefore (\ref{SDE_SUSY1})$\Leftrightarrow$(\ref{SDE_SUSY2}). Moreover, using 
Tr$(\sigma^m\sigma^n) = 2\eta^{mn}$ and $-2q.p=(q-p)^2-p^2-q^2$ we end up with (\ref{SDE_SUSY3})$\Leftrightarrow$(\ref{SDE_SUSY2}). This fact is obviously a consequence of supersymmetry. 
Finally, within this ansatz, we only have one equation to solve
\begin{equation}
 G(p^2) = 1 - \frac{2a}{\pi}\int\d^4l\frac{1}{\left[G(l^2)l^2+m^2\right](l+p)^2} + S
\end{equation}
with $S$ a counter-term and $a=\frac{g^2}{2\pi^3}$ the fine-structure constant of the theory. The Feynman integral above can be computed by performing the angular integral (\ref{angint}) as in the Yukawa model. We end up with
\begin{equation*}
 G(p^2) = 1 - \frac{2a}{p^2}\left(\int_0^{p}\frac{l^3\d l}{G(l^2)l^2+m^2} + p^2\int_{p}^{+\infty}\frac{l\d l}{G(l^2)l^2+m^2}\right) + S.
\end{equation*}
The counterterm $S$ is fixed by the initial condition $G(\mu^2)=1$. This provides:
\begin{equation*}
 S = \frac{2a}{\mu^2}\left(\int_0^{\mu}\frac{l^3\d l}{G(l^2)l^2+m^2} + \mu^2\int_{\mu}^{+\infty}\frac{l\d l}{G(l^2)l^2+m^2}\right).
\end{equation*}
Thus the Schwinger-Dyson equation simply becomes:
\begin{equation}
 G(p^2) = 1-2a\left(\frac{1}{p^2}\int_0^p\frac{l^3\d l}{G(l^2)l^2+m^2} - \frac{1}{\mu^2}\int_0^{\mu}\frac{l^3\d l}{G(l^2)l^2+m^2} - \int_{\mu}^p\frac{l\d l}{G(l^2)l^2+m^2}\right).
\end{equation}
Let us rewrite the previous equation with $x=p^2$ and $y=l^2$
\begin{equation} \label{SDE1}
 G(x) = 1 + a\int_{\mu^2}^x\frac{\d y}{G(y)y + m^2} + F(\mu^2) - F(x)
\end{equation}
with
\begin{equation}
 F(x) = a\int_0^x\frac{\d y}{G(y)y+m^2}\frac{y}{x}.
\end{equation}
Taking two derivatives with respect to $x$ of (\ref{SDE1}) it comes:
\begin{equation} \label{SDE2}
 \left[G(x) + \frac{m^2}{x}\right]\text{D}(\text{D}+1)G(x) = a
\end{equation}
with once again $D=x\frac{\d}{\d x}$. Now, we will see how, following the footsteps of \cite{BrKr99}, we can reach non-perturbative informations about our theories using the Schwinger--Dyson 
equations (\ref{SDEnSUSY2a}-\ref{SDEnSUSY2b}) and (\ref{SDE2}). We will start by studying the ultraviolet limit of the massive Yukawa model.

\section{Ultraviolet limit of the massive Yukawa model}

In this limit the exterior impulsion is much higher than the mass of the fermion. So the equations (\ref{SDEnSUSY2a}-\ref{SDEnSUSY2b}) become:
\begin{subequations}
 \begin{align}
  & G^2(x)D(D+1)M(x) = -aM(x) \label{2a} \\
  & G(x)D(D+2)G(x) = -a \label{2b}.
 \end{align}
\end{subequations}
We obviously start with the equation (\ref{2b}) since it is only an equation on $G$. It is actually the equation solved in \cite{BrKr99}, so we will not give all the details of the resolution.
Let us just say that the right change of variables is:
\begin{subequations}
 \begin{align*}
  & z = \left(\frac{x}{\mu^2}\right)^2 \\
  & \tilde{G} = \sqrt{\frac{2}{a}}zG(\mu^2\sqrt{z}).
\end{align*}
\end{subequations}
Then the equation (\ref{2b}) is simply:
\begin{equation*}
 2\tilde{G}(z)\tilde{G}''(z) = -1 \Leftrightarrow \left(\tilde{G}'(z)\right)^2 = \ln\tilde{G}(z) + \text{cste}.
\end{equation*}
So we can simply write $\tilde{G}$ as a function of $p:=\tilde{G}'$. Then looking for a differential equation for $\alpha=z/\tilde{G}$ and finding its asymptotic expansion allowed the authors 
of \cite{BrKr99} to find a simple parametric solution of (\ref{2b}):
\begin{subequations}
 \begin{align}
  & G(q^2) = \sqrt{\frac{a}{2\pi}}\frac{1}{\exp(p^2)\text{erfc}(p)} \label{solUV} \\
  & q^2 = \mu^2\sqrt{\frac{\text{erfc}(p)}{\text{erfc}(p_0)}}.
 \end{align}
\end{subequations}
Moreover the parameter $p_0 := \tilde{G}(1)$ is linked to the anomalous dimension $\delta$ by:
\begin{equation}
 p_0 = \frac{1}{\sqrt{2a}}(\gamma+2).
\end{equation}
The relation between the external impulsion $q^2$ and the parameter $p$ was found by using the initial conditions on $\alpha$ and the intermediate result $\tilde{G} = \exp(p^2-p_0^2)$. A 
more detailed derivation is given in section 5 for a similar problem. To tackle (\ref{2a}) let us define the analog of $\tilde{G}$ for the mass function
\begin{equation*}
 \tilde{M}(z) = \sqrt{z}M(\mu^2\sqrt{z}).
\end{equation*}
Then, (\ref{2a}) written in term of $\tilde{M}$ and $\tilde{G}$ is (after simplifications):
\begin{equation}
 \frac{\tilde{G}^2(z)}{\sqrt{z}}\frac{2}{\tilde{M}(z)}\frac{\d}{\d z}\left[\sqrt{z}\tilde{M}'(z)\right] = -1.
\end{equation}
We can rewrite this equation as a differential equation in $p$ by using the resolution of (\ref{2b}) and $\frac{\d p}{\d z}=\tilde{G}''=-\frac{1}{2\tilde{G}}$. To do so, we will write $h(p)$ the 
function defined as $\tilde{M}$ seen as a function of $p$: $h(p):=\tilde{M}(z(p))$. We end up with:
\begin{equation} \label{eqh}
 h''(p) + [2p+f'(p)]h'(p)+2h(p) = 0.
\end{equation}
Here the prime stands for a derivative with respect to $p$, while in the previous equations it was for a derivative with respect to $z$ and $f(p)$ is  by definition:
\begin{equation}
 f(p) = \frac{1}{2}\ln\left[\text{erfc}(p)\right] \Rightarrow f'(p) = -\frac{\tilde{G}(z(p))}{z(p)}.
\end{equation}
The last equality comes from the solution (\ref{solUV}) and the definition of $\tilde{G}$.
Notice that the complementary error function erfc is defined as
\begin{equation*}
 \text{erfc}(z) = \frac{2}{\sqrt{\pi}}\int_z^{+\infty}e^{-t^2}\d t.
\end{equation*}
Therefore it is a positive function and $f$ is a well-defined real function. We solve (\ref{eqh}) with one last change of variable: let us define the new function $B(p)$:
\begin{equation}
 h(p) = e^{B(p)}.
\end{equation}
From the definition of $h(p)$ and since $M(q^2)$ is the mass function and therefore is expected to be positive, we see that $B(p)$ has to be a real function. Now the equation (\ref{eqh}) is a first 
order differential equation in $B'(p)$:
\begin{equation*}
 B''(p) + \left[f'(p)+2p+1\right]B'(p) + 2 = 0
\end{equation*}
Whose solution is:
\begin{equation} \label{Bprime}
 B'(p) = e^{-p^2-p-f(p)}\left(\lambda - 2\int_{p_0}^pe^{f(x)+x+x^2}\d x\right)
\end{equation}
We can determine the integration constant $\lambda$ by using the initial conditions.
\begin{align*}
 B'(p_0) = -\sqrt{\frac{2}{a}}(1+\delta) \Leftrightarrow 
\begin{cases}
 & h(p_0)=1 \\
 & h'(p_0) = -\sqrt{2}{a}(1+\delta)
\end{cases}
\end{align*}
With $\delta := q^2\frac{\d M(q^2)}{\d q^2}|_{q^2=\mu^2}$ the massive anomalous dimension. In order to have a more readable text, the obtained $\lambda$ will only be written below, in the 
final solution. The next step is to integrate (\ref{Bprime}):
\begin{equation}
 B(p) = \nu + \int_{p_0}^p\text{erfc}(t)^{-1/2}e^{-t^2-t}\left(\lambda - 2\int_{p_0}^t\text{erfc}(s)^{1/2}e^{s^2+s}\d s\right)\d t
\end{equation}
And $\nu$ is very simply determined by $h(p_0) := \tilde{M}(1) = 1 \Leftrightarrow B(p_0)=0 \Leftrightarrow \nu = 0$. Let us notice that $B(p)$ is real as expected. Hence we get $\tilde{M}$ and 
thus $M$. Putting everything together we end up with a parametric solution for the massive Yukawa model in the ultraviolet limit:
\begin{subequations}
 \begin{align}
  & G(q^2) = \sqrt{\frac{a}{2\pi}}\frac{1}{\exp(p^2)\text{erfc}(p)} \\
  & M(q^2) = \sqrt{\frac{\text{erfc}(p_0)}{\text{erfc}(p)}}\exp\left(\int_{p_0}^p\text{erfc}(t)^{-1/2}e^{-t^2-t}\left[\lambda - 2\int_{p_0}^t\text{erfc}(s)^{1/2}e^{s^2+s}\d s\right]\d t\right) \\
  & \lambda = -\sqrt{\frac{2}{a}}(1+\delta)e^{p_0^2+p_0}\text{erfc}(p_0)^{1/2} \\
  & q^2 = \mu^2\sqrt{\frac{\text{erfc(p)}}{\text{erfc}(p_0)}}
 \end{align}
\end{subequations}
Now we can move on to the other limit: the infrared one.

\section{Infrared limit of the massive Yukawa model}

In this case we have $x<<m^2$. However, two possibilities have to be separated. Indeed either $a$, the coupling constant of the theory is big enough so the RHS of 
(\ref{SDEnSUSY2a}-\ref{SDEnSUSY2b}) is of the order of the LHS, either it is not. Let us start with the first case, that is:
\begin{equation} \label{hypa}
 a \sim \frac{m^2}{x}
\end{equation}
This case is called ``soft infrared'' since the external momenta is not small enough to have the coupling constant negligible. Then the equations (\ref{SDEnSUSY2a}-\ref{SDEnSUSY2b}) become:
\begin{subequations}
 \begin{align} 
  & m^2M(x)D(D+1)M(x) = ax  \label{16a} \\
  & m^2M^2(x)D(D+2)G(x) = axG(x) \label{16b}.
\end{align}
\end{subequations}
We will start by solving (\ref{16a}) since there is only one unknown function in it. Using the reduced coupling constant $\tilde{a} := a\frac{\mu^2}{m^2}$ let us change the variables:
\begin{subequations}
 \begin{align}
  & z = \left(\frac{x}{\mu^2}\right)^2 \\
  & \tilde{M}(z) = \sqrt{\frac{z}{\tilde{a}}}M(\mu^2\sqrt{z}).
\end{align}
\end{subequations}
Then (\ref{16a}) becomes:
\begin{equation} \label{eqMtilde}
 4\tilde{M}(z)\frac{\d}{\d z}\left[\sqrt{z}\frac{\d}{\d z}\tilde{M}(z)\right] = 1.
\end{equation}
The previous analysis can be done one more time to this equation, but a critical step will there be missing because of the $\sqrt{z}$ into the outer derivative: we cannot integrate the 
equation. Actually, there is no definition of $z$ and $\tilde{M}$ that would makes the new version of (\ref{eqMtilde}) integrable. Fortunately we are saved by noticing that it exists a simple 
solution to (\ref{eqMtilde}):
\begin{equation} \label{solM}
 \tilde{M}(z) = \frac{2}{\sqrt{3}}z^{3/4}.
\end{equation}
This solution satisfies the initial condition $\tilde{M}(1)=1/\sqrt{\tilde{a}}$ if, and only if, the reference scale is
\begin{equation}
 \mu = \frac{m}{2}\sqrt{\frac{3}{a}}.
\end{equation}
This is coherent with the hypothesis (\ref{hypa}) since the the impulsion $q$ should be of the same order than the impulsion of reference. Notice that, at this impulsion of reference, one gets 
$\tilde{a} = 3/4$. Now, plugging the solution (\ref{solM}) into the equation (\ref{16b}) one gets, for the function $\tilde{G}(z) = zG(\mu^2\sqrt{z})$, the very simple equation:
\begin{equation}
 \tilde{G}''(z) = \frac{3}{16}\tilde{G}(z).
\end{equation}
Which has the simple solution
\begin{equation*}
 \tilde{G}(z) = A\exp\left(\frac{\sqrt{3}}{4}z\right) + B\exp\left(-\frac{\sqrt{3}}{4}z\right).
\end{equation*}
The coefficients $A$ and $B$ are determined, thanks to the initial conditions $\tilde{G}(1)=1$ and $\tilde{G}'(1)=1+\gamma/2$. To summarize, we get a solution to the Schwinger-Dyson equation to 
the massive Yukawa model with a coupling constant of the order of $\frac{m^2}{q^2}$.
\begin{subequations}
 \begin{align}
& M(q^2) = \frac{3}{a}\frac{m^2}{q^2} \\
& G(q^2) = \frac{9}{16a^2}\frac{m^4}{q^4}\frac{1}{2\sqrt{3}}\left[(\sqrt{3}+4+2\gamma)\exp\left(\frac{\sqrt{3}}{4}\left[\frac{16a^2q^4}{9m^4}-1\right]\right)+(\sqrt{3}-4-2\gamma)\exp\left(-\frac{\sqrt{3}}{4}\left[\frac{16a^2q^4}{9m^4}-1\right]\right)\right]
 \end{align}
\end{subequations}
Now, let us look at the case for $a$ is not big enough to cancel the fact that we are in the infrared regime. By opposition to the previous case, this one is called ``deep infrared''. This case 
is more interesting from a physical point of view since the small energies (at which $a<<\frac{m^2}{q^2}$) are easier to reach. Moreover, it is at those low energies that the coupling constant 
of QCD becomes too big to allow perturbative computations\footnote{i.e. of order of unit, which is still much lower than the ratio $m^2/q^2$.} making the Schwinger--Dyson equations of interest 
for physicists as a door to non-perturbative regimes. In this case, the RHS of equations (\ref{SDEnSUSY2a}-\ref{SDEnSUSY2b}) is negligible, so this system becomes:
\begin{align*}
  & m^2M^2(x)D(D+1)M(x) = 0 \\
  &  m^2M^2(x)D(D+2)G(x) = 0.
\end{align*}
Since we are not looking for vanishing solutions, this system actually decouples:
\begin{subequations}
 \begin{align}
 & D(D+1)M(x) = 0 \\
 & D(D+2)G(x) = 0.
\end{align}
\end{subequations}
And those are very easy to solve with the following initial conditions:
\begin{subequations}
 \begin{align}
  & M(\mu^2) = G(\mu^2) = 1 \\
  & q^2\frac{\d M(q^2)}{\d q^2}|_{q^2=\mu^2} = \delta \\
  & q^2\frac{\d G(q^2)}{\d q^2}|_{q^2=\mu^2} = \gamma.
\end{align}
\end{subequations}
Then one ends up with:
\begin{subequations}
 \begin{align}
  & M(q^2) = 1+\delta-\frac{\delta\mu^2}{q^2} \\
  & G(q^2) = 1+\frac{\gamma}{2}-\frac{\gamma}{2}\frac{\mu^4}{q^4}
 \end{align}
\end{subequations}
This solution being obviously for every scale of reference $\mu$.

\section{Solution of the supersymmetric model}

Now, let us look for a solution to the Schwinger--Dyson equation (\ref{SDE2}). First, we switch to dimensionless variables, and define the new parameters and functions:
\begin{subequations}
 \begin{align}
  & z = \frac{x}{\mu^2} \\
  & \tilde{m}^2 = \frac{m^2}{\mu^2}\frac{1}{\sqrt{2a}} \\
  & \tilde{G}(z) = \frac{1}{\sqrt{2a}}zG(\mu^2z).
\end{align}
\end{subequations}
Then the equation (\ref{SDE2}) becomes
\begin{equation}
 2\left[\tilde{G}(z)+\tilde{m}^2\right]\tilde{G}''(z) = 1. \label{eqGtildeSUSY}
\end{equation}
Let us notice than $\tilde{m}$ is still constant with respect to $z$ since we take $\mu^2$ being fixed. Then the above equation can easily be integrated to:
\begin{equation*}
 \left(\tilde{G}'(z)\right)^2 = \ln\left(\tilde{G}(z)+\tilde{m}^2\right)+\text{cste}.
\end{equation*}
Now, let us use the parameter $p=\tilde{G}'(z)$. We can write $\tilde{G}(z)$ as a function of $p$:
\begin{equation} \label{formeG}
 \tilde{G}(z) = \left(\tilde{m}^2+\frac{1}{\sqrt{2a}}\right)\exp(p^2-p_0^2)-\tilde{m}^2
\end{equation}
with $p_0:=\tilde{G}'(1)$ and where we have used the initial condition $\tilde{G}(1)=\frac{1}{\sqrt{2a}} \Leftrightarrow G(\mu^2)=1$. We can then determine the parametric solution by 
studying:
\begin{equation} \label{alpha}
 \alpha = \frac{z}{2(\tilde{G}+\tilde{m}^2)}.
\end{equation}
Indeed, using the definition of $p$ and the equation (\ref{formeG}) we obtain a differential equation for $\alpha$
\begin{equation}
 \alpha = \frac{1}{2p}-\frac{1}{2p}\frac{\d\alpha}{\d p}.
\end{equation}
Developing the solution at infinity gives the asymptotic expansion:
\begin{equation} \label{expansionAlpha}
 \alpha(p) \simeq \frac{1}{2p}+\frac{1}{2p}\sum_{n=1}^{+\infty}\frac{(2n-1)!!}{(2p^2)^n}.
\end{equation}
Now, using the definition (\ref{alpha}) of $\alpha$ and the one of the anomalous dimension
\begin{equation*}
 \gamma = q^2\frac{\d G(q^2)}{\d q^2}|_{q^2=\mu^2},
\end{equation*}
we get the relations:
\begin{subequations}
 \begin{align} 
  & \alpha(p_0) = \sqrt{\frac{a}{2}}\frac{1}{\frac{m^2}{\mu^2}+1} \label{refInita} \\
  & p_0 = \frac{1}{\sqrt{2a}}(1+\gamma) \label{refInitb}.
\end{align}
\end{subequations}
Such relations were also found in \cite{BrKr99}. The second of the above relations comes from
\begin{equation*}
 p_0:=\frac{1}{\sqrt{2a}}\frac{\d}{\d z}(zG)|_{z=1}.
\end{equation*}
Then, we easily get the parametric solution by noticing that the expansion (\ref{expansionAlpha}) is the one of:
\begin{eqnarray*}
 \alpha(p) & = & \frac{1}{2}\sqrt{\pi}e^{-p^2}\Re[\text{erfi}(p)] \\
           & = & \frac{1}{2}\left(\sqrt{\pi}e^{-p^2}\text{erfi}(p) + \text{i}\sqrt{\pi}e^{-p^2}\right) \\
           & = & \frac{\text{i}}{2}\sqrt{\pi}e^{-p^2}\text{erfc}(\text{i}p)
\end{eqnarray*}
With erfc the complementary error function and erfi the imaginary error function. In the following, we will simply write $\E(p):=\Re[\text{erfi}(p)]$ to emphasize that it is a real function. 
Hence we have:
\begin{equation} \label{formAlpha}
 \alpha(p) = \frac{1}{2}\sqrt{\pi}e^{-p^2}\E(p).
\end{equation}
Now, we have everything to write the parametric solution to the equation (\ref{SDE2}). First, we need to write $z$ as a function of $p$ (since we will write $G$ as a function of $p$). Using its 
definition and the form (\ref{formeG}) of $\tilde{G}$ we get
\begin{equation*}
 1 = 2\exp(p^2-p_0^2)\left(\tilde{m}+\frac{1}{\sqrt{2a}}\right)\frac{\d p}{\d z}.
\end{equation*}
As in the non-supersymmetric case, using the equation (\ref{eqGtildeSUSY}) and the definition (\ref{alpha}) of $\alpha$ we obtained $\frac{\d p}{\d z}=\frac{\alpha}{z}$. Hence, using the above 
formula (\ref{formAlpha}), we end up with
\begin{equation*}
 z = \sqrt{\pi}\E(p)\left(\tilde{m}+\frac{1}{\sqrt{2a}}\right)\exp(-p_0^2).
\end{equation*}
This could be written on a much simpler form by using the relation (\ref{refInita}) together with (\ref{formAlpha}). Then one obtains for $q^2$ the form written below, in the  complete solution. 
Finally, writing $G(q^2)=\sqrt{2a}\left(\frac{1}{2\alpha}-\frac{\tilde{m}^2}{z}\right)$ we get the parametric solution to the Schwinger-Dyson equation of our massive supersymmetric model:
\begin{subequations}
 \begin{align}
  & G(q^2) = \sqrt{\frac{2a}{\pi}}\frac{e^{p^2}}{\E(p)}-\frac{m^2}{\mu^2}\frac{\E(p_0)}{\E(p)} \\
  & q^2 = \mu^2\frac{\E(p)}{\E(p_0)}.
\end{align}
\end{subequations}
To our knowledge, this is the second (the first being in \cite{BrKr99}) known exact solution of a Schwinger--Dyson equation, and the first for a theory with a mass term.

\section*{Conclusion}

Using kinematic renormalization, we have been able to write down the renormalized Schwinger--Dyson equations of the massive Yukawa model, and of a massive linear version of a Wess--Zumino 
model. The non-supersymmetric model has been studied both in ultraviolet (UV) and infrared (IR) limit. In the UV limit, the equation fulfilled by the wavefunction is the same as the one 
solved in \cite{BrKr99}. We could then use the result of this paper to determine the mass function of the Yukawa model. For the IR limit, two cases had to be separated, called soft 
and deep IR. In the soft case, an explicit solution to the Schwinger--Dyson equation has been found at a given impulsion of reference. For the deep IR case, an explicit solution has been found as 
well, but for every impulsion of reference.

For the supersymmetric massive model, supersymmetry could be used twice to simplify the problem: it provides a non-renormalization theorem and allows to take all the wavefunctions to be equal. 
We end up with a single differential equation instead of a system of coupled differential equations. Then, using a series of transformations, a parametric solution has been found without taking 
any limit. It is, as far as we know, the first exact solution of a Schwinger--Dyson equation (since the system reduces to only one equation) with a mass term. Therefore, we hope to have proven 
that the method of \cite{BrKr99} is worth further investigations.

There are many tracks that one might follow to go beyond our analysis. The most natural question to ask is whether one could find the perturbations to the solutions of the Yukawa model in the UV 
and IR limit. Although this is not a trivial task due to the complexity of some of our functions, it could still be tackled. When doing so, very interesting questions of convergence of the 
solutions arise, that are currently under investigation.

Another approach to solve the Schwinger--Dyson equations of the massive Yukawa theory beyond the UV and IR limit could be the use of the technology of special functions. Important progresses have
been made (and are made) to use them for evaluating Feynman integrals. One can look at \cite{Pa14} for a nice survey of the field. Since those functions obey some functional equations, we could 
study Schwinger--Dyson equations in their light.

A maybe simpler task would be to numerically compare the UV and IR solutions of the Yukawa model and try to fit them into a solution covering the full range of possible values of the external 
impulsion. Instead of looking for perturbations of the solutions found in parts $3$ and $4$, one could look for pertubations of the equations solved in those parts. Indeed the solutions are 
rather complicated and not well-suited for numerical investigations. Therefore it would be much more convenient to deal with modified equations to unravel the behaviour of the massive Yukawa 
model beyond the UV and IR limits.

Moreover, on the supersymmetric side, a numerical study of the anomalous dimension of our supersymmetric model had been made in \cite{BeLoSc07} (with $m=0$) using a Pad\'e--Borel resummation 
method. It would be very interesting to compare this study to our in order to gain a better understanding of the effect of the mass on the solution. 

Even for non-linear Schwinger--Dyson equation, it as been shown in \cite{BeCl14} that the asymptotic behaviour of the solution of the Schwinger--Dyson equation (in this case written for the 
anomalous dimension) is given by a term which is the linear part of the full non-linear Schwinger--Dyson equation. Thus our results could be used in more physical cases, at least to reach 
informations about the asymptotic behaviour of the solution seen as a series of the coupling constant of the theory.

\section*{Acknowledgements}

I am very thankful to Marc P. Bellon for introducing me to the fascinating subject of Schwinger--Dyson equations and having suggested this particular project to me. I would also like to thank 
Lucien F. F. Heurtier for discussions and helpful reading suggestions concerning non-renormalization theorems and SUSY breaking.

\bibliographystyle{unsrt}
\bibliography{renorm}

\begin{thebibliography}{10}

\bibitem{MeMoRoSa14}
C.~Mezrag, H.~Moutarde, J.~Rodrigues-Quintero, and F.~Sabati\'e.
\newblock {T}oward a {P}ion {G}eneralized {P}arton {D}istribution {M}odel
  {F}rom {D}yson--{S}chwinger equations.
\newblock 2014.

\bibitem{KrYe2006}
Dirk Kreimer and Karen Yeats.
\newblock {An etude in non-linear Dyson-Schwinger equations}.
\newblock {\em Nucl. Phys. Proc. Suppl.}, 160:116--121, 2006.

\bibitem{BeSc12}
Marc Bellon and Fidel~A. Schaposnik.
\newblock Higher loop corrections to a {S}chwinger--{D}yson equation.
\newblock {\em Lett.\ Math.\ Phys.}, 103:881--893, 2013.

\bibitem{BeCl13}
Marc~P. Bellon and Pierre~J. Clavier.
\newblock Higher order corrections to the asymptotic perturbative solution of a
  {S}winger--{D}yson equation.
\newblock {\em Lett.\ Math.\ Phys.}, 104:1--22, 2014.

\bibitem{BeCl14}
Marc~P. Bellon and Pierre~J. Clavier.
\newblock Study of a {S}chwinger--{D}yson {E}quation in the {B}orel {P}lane.
\newblock {\em To be published}, 2014.

\bibitem{BrKr99}
D.~J. Broadhurst and D.~Kreimer.
\newblock Exact solutions of {D}yson--{S}chwinger equations for iterated
  one-loop integrals and propagator-coupling duality.
\newblock {\em Nucl.\ Phys.}, B 600:403--422, 2001.

\bibitem{Itzykson}
C.~Itzykson and J.B. Zuber.
\newblock {\em {Q}uantum {F}ield {T}heory}.
\newblock 1978.

\bibitem{BeLoSc07}
{M}arc {B}ellon, {G}ustavo {L}ozano, and {F}idel {S}chaposnik.
\newblock {H}igher loop renormalization of a supersymmetric field theory.
\newblock {\em {P}hysics {L}etters {B}}, 650:293--297, 2007.

\bibitem{Se93}
N.~Seiberg.
\newblock {N}aturalness {V}ersus {S}upersymmetric {N}on-renormalization
  {T}heorems.
\newblock {\em {P}hys. {L}ett. {B}}, 318:469--475, 1993.

\bibitem{De89}
Jean-Pierre Derendinger.
\newblock {L}ecture notes on globally supersymmetric theories in
  four-dimensions and two-dimensions.
\newblock 1989.

\bibitem{Pa14}
Erik Panzer.
\newblock Feyman integrals via hyperlogarithms.
\newblock {\em Loops and Legs in Quantum Field Theory}, 2014.

\end{thebibliography}

\end{document}